\documentclass[%
 preprint,
 amsmath,amssymb,
 aps,
]{revtex4-1}

\usepackage{graphicx}
\usepackage{dcolumn}
\usepackage{bm}
\usepackage{amsfonts}
\usepackage{hyperref}
\usepackage{epstopdf}

\newcommand{\snn}{\sqrt{s_\text{NN}}}

\begin{document}

\preprint{}

\title{Particle decay from statistical thermal model in high energy nucleus-nucleus collision}

\author{Ning Yu}
\thanks{ning.yuchina@gmail.com}
\affiliation{School of Physics \& Electronic Engineering, Xinyang Normal University, Xinyang 464000, China}
\author{Xiaofeng Luo}
\thanks{xfluo@mail.ccnu.edu.cn}
\affiliation{Key Laboratory of Quark \& Lepton Physics (MOE) and Institute of Particle Physics, \\Central China Normal University, Wuhan, 430079, China.}

\date{\today}

\begin{abstract}
In high energy nucleus-nucleus collisions, it is difficult to measure the contributions of resonance strong decay and weak decay to the final measured hadrons as well as the corresponding effects on some physical observables. To provide a reference from statistical thermal model, we performed a systematic analysis for the energy dependence of particle yield and yield ratios in Au + Au collisions. We found that the primary fraction of final hadrons decreases with increasing collision energy and somehow saturates around $\snn$ = 10 GeV, indicating a limiting temperature in hadronic interactions. The fraction of strong or weak decay for final hadrons show a different energy dependence behavior comparing to the primarily produced hadrons. These energy dependences of various particle yield and yield ratios from strong or weak decay can provide us with baselines for many hadronic observables in high energy nucleus-nucleus collisions.
\begin{description}\item[PACS numbers]
\verb+25.75.Nq, 24.10.Lx, 24.10.Pa+
\end{description}
\end{abstract}

\maketitle


Ultra-relativistic nucleus-nucleus collisions can create a new state of matter, the quark-gluon plasma (QGP) in laboratories. The phase structure of strong interactions, where quarks and gluons are deconfined, can be studied by quantum chromodynamics (QCD). After this strongly coupled QGP was observed at the Relativistic heavy-ion Collider (RHIC)~\cite{ADAMS2005102}, attempts are being made to vary the colliding beam energy and to research the thermodynamics properties of QCD matter expressed in terms of a temperature vs baryo-chemical potential ($T_{\rm {ch}} -\mu_B$) phase diagram, which lies at the heart of what the RHIC Beam Energy Scan (BES) program is all about~\cite{Stephanov,MOHANTY2009899c,LUO201675}. 

Thermodynamic properties of the QCD phase diagram can be decoded via analysis of particle production in heavy-ion collisions. This is achieved by using a statistical thermal model, in which the yields of different particle species contain characteristic features to determine the $T_{\rm {ch}}$ and $\mu_B$ of the system at freeze-out. During the system evolution, there are two types of freeze-out: chemical and kinetic freeze-out. Chemical freeze-out is typically supposed to happen when inelastic scattering stops, and the particle identities are fixed until they decay~\cite{Vogt}. After chemical freeze-out, elastic interactions among the particles are still ongoing which leads to changes in the momentum of the particles. When the average inter-particle distance becomes large enough to make the elastic interactions stop, the system is said to have reached kinetic freeze-out. At this stage, the transverse momentum spectra of the produced particles becomes fixed. It is a surprising success that statistical thermal model can reproduce essential features of particle production in high energy nuclear collisions~\cite{PhysRevC.65.027901,PhysRevC.73.044905,ANDRONIC2007334,PhysRevC.73.034905}, suggesting that statistical production is a general property of the hadronization process.

It must be noted that the chemical freeze-out, i.e. from hadrons and especially, hadronic resonances, happens before they decay, including strong and electromagnetic decays of high-mass resonance, and weak decay from heavy flavor hadrons. In experiment, contamination of hadrons from strong decay $h_s$ is difficult to measure due to their short lifetime. Contamination of hadrons from weak decay $h_w$ can be extracted by DCA(distance of closest approach) distribution~\cite{PhysRevC.88.044910}. The fraction of $h_s$ or $h_w$ for final hadrons shows different energy dependence behavior comparing to the primarily produced hadrons $h_p$, which are not real physical signals we care about. During the first phase of the RHIC BES (2010–2014), the STAR experiment has measured the collision energy dependence of many observables, such as the cumulants of net-proton, net-charge and net-kaon multiplicity distribution~\cite{PhysRevLett.105.022302,PhysRevLett.112.032302,PhysRevLett.113.092301,2018551,Luo:2015ewa,Luo:2017faz}, the directed flow $dv_1/dy$ for net-protons~\cite{PhysRevLett.112.162301}, and coalescence parameters for deuterons~\cite{YU2017788,Yu:2018kvh}. Non-monotonic behaviors were found in the energy dependence of these observables. In order to provide a model reference to the decay effect and find which observables are sensitive to the decay effect, a statistic thermal model will be used to study the energy dependence of $h_s$ and $h_w$. At the stage of chemical freeze-out, the particle abundance of species $i$ can be parametrized by

\begin{equation}\label{nthermal}
\frac{N_i}{V}=\frac{g_i}{2\pi^2}\sum_{k=1}^{\infty}(\mp)^{k+1}\frac{m_i^2T_{\rm {ch}}}{k}K_2\left(\frac{km_i}{T_{\rm {ch}}}\right)e^{k\mu_i/T_{\rm {ch}}},
\end{equation}
where
\begin{equation}
\mu_i=\mu_BB_i+\mu_QQ_i+\mu_SS_i,
\end{equation}
and $g_i$ is the spin-iso-spin degeneracy factor; $T_{\rm {ch}}$ is the chemical freeze-out temperature;  $B_i$, $S_i$, $Q_i$ are the baryon number, strangeness, and charge, respectively, of hadron species $i$; $\mu_B$, $\mu_S$, and $\mu_Q$ are the corresponding chemical potentials for these conserved quantum numbers. The code THERMUS~\cite{WHEATON200984} is utilized to perform a thermal calculation of particle yields. Within the model, there is a freedom regarding the ensemble with which to treat conserved numbers $B$, $S$, and $Q$ in strong interactions. The chemical potentials for each of these quantum numbers allow fluctuations about conserved averages, which is a reasonable approximation only when the number of particles carrying the quantum number concerned is large. Three ensembles can be employed in the model. Those are the grand-canonical ensemble (GCE), canonical ensemble (CE), and mix-strangeness canonical ensemble (SCE). The GCE is the most widely used in the application to heavy-ion collisions. In GCE, the Boltzmann approximation ($k = 1$ in Eq. ~\ref{nthermal})  is reasonable for all particles except the pions,

\begin{equation}
N_{i}=\frac{g_iV}{2\pi^2}m_i^2TK_2\left(\frac{m_i}{T_{\rm {ch}}}\right)e^{\beta\mu_i}.
\end{equation}
For this approximation analysis, the deviation of quantum statistical effect for pions is at the level of 10\%, while, for kaons, the deviation peaks at between 1 and 2\%. For all other mesons, the deviation is less than the 1\% level. For baryons, the deviation is extremely small.

The energy dependence of chemical freeze-out parameters $T_{\rm {ch}}$ and $\mu_B$ are obtained from statistical hadronization analysis of hadron yields~\cite{Andronic:2017pug},

\begin{eqnarray}
T_{\rm {ch}}&=&\frac{T_{\rm {ch}}^{\lim}}{1+\exp\left(2.60-\ln(\snn)/0.45\right)}\\
\mu_B&=&\frac{\mu_B^{\lim}}{1+0.288\snn},
\end{eqnarray}
where $T_{\rm {ch}}^{\lim}=158.4\pm 1.4$ MeV, $\mu_B^{\lim}=1307.5$ MeV. With these thermal parameters on energy, we can get the energy dependence of $h_p$, $h_s$, and $h_w$.

\begin{figure}
\resizebox{1\columnwidth}{!}{\includegraphics{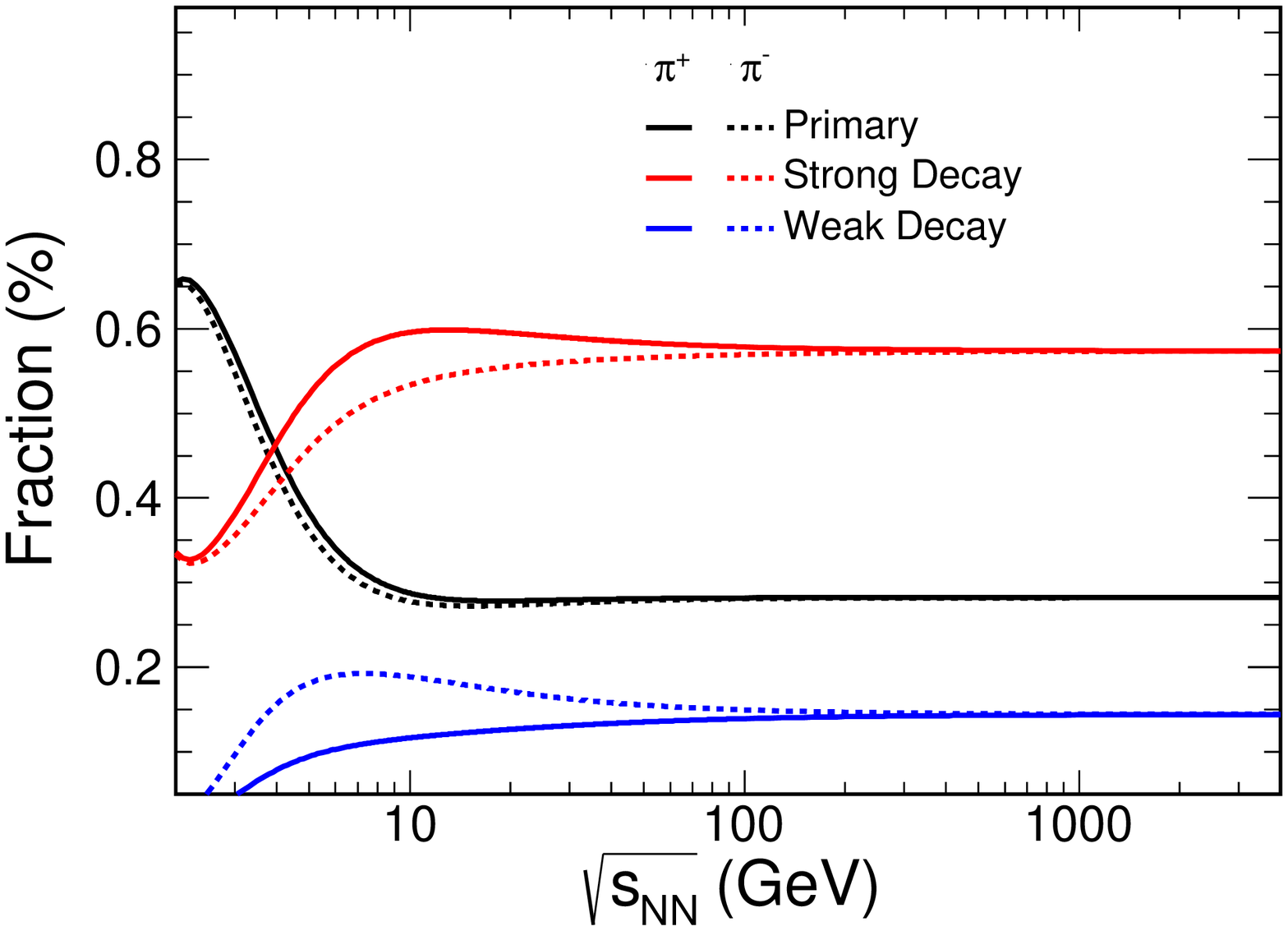}}
\caption{Energy dependence of $\pi^{+}$ and $\pi^{-}$ fractions for primary production, strong decay from high-mass resonance, and weak decay from heavy flavor hadrons.}
\label{fpi}
\end{figure}

\begin{figure}
\resizebox{1\columnwidth}{!}{\includegraphics{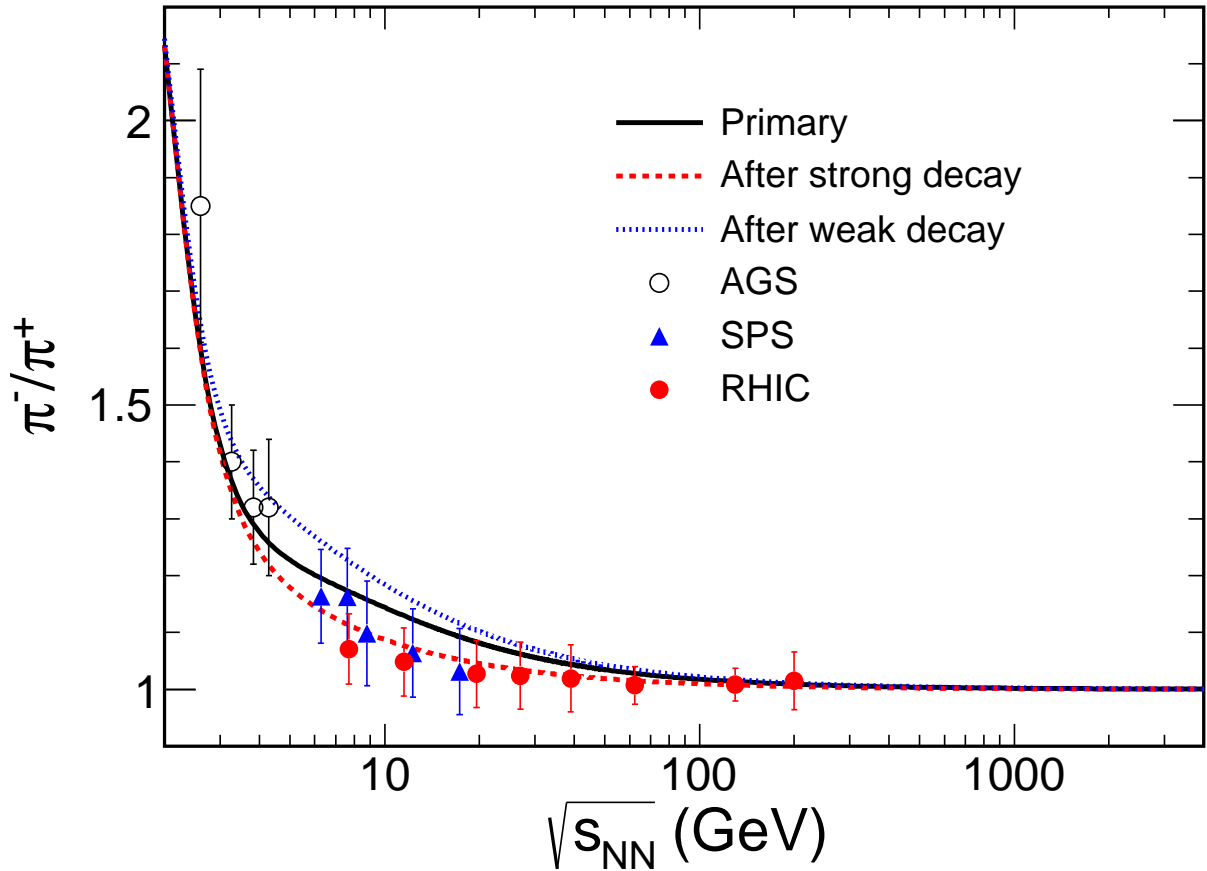}}
\caption{Energy dependence of particle ratios $\pi^{-}/\pi^{+}$ from the stage at primary production, after strong decay from high-mass resonance, and after weak decay from heavy flavor hadrons. Experimental results from AGS~\cite{AKIBA1996139,PhysRevC.57.R466,20001,PhysRevC.62.024901,PhysRevC.60.064901,200053,PhysRevLett.88.102301}, SPS~\cite{PhysRevC.77.024903,PhysRevC.73.044910,PhysRevC.69.024902,PhysRevC.66.054902}, and RHIC~\cite{PhysRevC.96.044904,PhysRevC.81.024911,PhysRevC.79.034909,PhysRevLett.92.112301} of the most central collision are shown for comparison.}
\label{rpi}
\end{figure}

Figure~\ref{fpi} shows the energy dependence of $\pi^{+}$ and $\pi^{-}$ fractions for primary production, strong decay from high-mass resonance, and weak decay from heavy flavor hadrons, respectively. Those are $h_p/(h_p+h_s+h_w)$, $h_s/(h_p+h_s+h_w)$, and $h_w/(h_p+h_s+h_w)$. It can be found that energy dependence of these factions are different for $\pi^{+}$ and $\pi^{-}$. Fractions of $h_p$ for $\pi^{+}$ and $\pi^{-}$ are almost the same, they decrease from 65\% to 28\% with increasing collision energy. In the Boltzmann approximation, the particle ratio $\pi^{-}/\pi^{+}$ is related to the iso-spin effect as
\begin{equation}\label{iso}
\frac{h_p(\pi^{-})}{h_p(\pi^{+})}=\exp{\frac{-2\mu_Q}{T_{\rm {ch}}}}.
\end{equation}
The deviation of this approximation from Bose-Einstein is less than 5\% with $T_{\rm {ch}} <$ 180 MeV and $\mu_Q/T_{\rm {ch}} > -0.4$.  In nucleus-nucleus collision, $h_p(\pi^{-})/h_p(\pi^{+})$ is greater than 1 and decreases with $\snn$ and saturates to 1, which can be found in figure~\ref{rpi}. That is due to the fact that $\mu_Q$ is less than 0 and increases to zero with increasing $\snn$. The $h_s(\pi^{\pm})$ are mainly from $\Delta$ resonances at low collision energies. Ratio of pions from $\Delta$ decay can be calculated by

\begin{equation}
\frac{h_s(\pi^{-}\leftarrow\bar{\Delta})}{h_s(\pi^{+}\leftarrow\Delta)}=\exp{\frac{-2c_{\Delta}\mu_Q-2\mu_B}{T_{\rm {ch}}}}\approx\exp{\frac{-2\mu_B}{T_{\rm {ch}}}}<1,
\end{equation}
where $c_{\Delta} > 1$ is the effective charge of strong decay which contains the contribution of multi-charged $\Delta$. The contribution from $\Delta$ for $\pi^{-}$ is smaller than that for $\pi^{+}$. The decayed pions from short lived mesons, such as $\eta$, $\rho$ become significant with increasing energy, which gives the same contribution to the yields of $\pi^{+}$ and $\pi^{-}$. With these two kinds of strong decay, the fraction of $h_p(\pi^{-})$ is smaller than that of $h_p(\pi^{+})$ and the particle ratio $\pi^{-}/\pi^{+}$ is suppressed after strong decay. As a result, we cannot use Eq.~\ref{iso} with the ratio corrected by weak decay to extract iso-spin effect in nucleus-nucleus collision, which will underestimate real iso-spin effect. Components of $h_s(\pi^{\pm})$ increase with energy and saturate at the value of 57\% around $\snn = $10 GeV. $h_w(\pi^{\pm})$ are mainly from the channels below

\begin{eqnarray}
K_S^0&\rightarrow &\pi^{+}+\pi^{-}\nonumber\quad\quad\textrm{B.R.}=69.2\%\\
\Lambda(\bar{\Lambda})&\rightarrow &p(\bar{p})+\pi^{\mp}\nonumber\quad\quad\textrm{B.R.}=63.9\%\\
\Sigma^{+}(\bar{\Sigma}^{-})&\rightarrow &n(\bar{n})+\pi^{\pm}\nonumber\quad\quad\textrm{B.R.}=48.31\%\\
\Sigma^{-}(\bar{\Sigma}^{+})&\rightarrow &n(\bar{n})+\pi^{\mp}\nonumber\quad\quad\textrm{B.R.}=99.85\%
\end{eqnarray}

The ratio of pions from weak decay can be calculated as
\begin{eqnarray}
\frac{h_w(\pi^{-}\leftarrow\Lambda)}{h_w(\pi^{+}\leftarrow\bar{\Lambda})}&=&\exp{\frac{2\mu_B-2\mu_S}{T_{\rm {ch}}}}>1\\
\frac{h_w(\pi^{-}\leftarrow\bar{\Sigma}^{-}/\Sigma^{-})}{h_w(\pi^{+}\leftarrow\Sigma^{+}/\bar{\Sigma}^{-})}&\approx&1+\frac{51.54\% \times \left(\exp{\frac{2\mu_B-2\mu_S}{T_{\rm {ch}}}}-1\right)}{48.31\%\times\exp{\frac{2\mu_B-2\mu_S}{T_{\rm {ch}}}}+99.85\%}>1,
\end{eqnarray}
the value 51.54\% is from the difference of branch ratio between $\Sigma^{+}$ and $\Sigma^{-}$ decay to pions. The same yield of $\pi^{+}$ and $\pi^{-}$ is created in $K_S^0$ weak decay. So, more $\pi^{-}$s are created in weak decay than $\pi^{+}$s especially at low energy, which will enhance the $\pi^{-}/\pi^{+}$ after weak decay. The $h_w(\pi^{+})$ reach the maximum around $\snn = $8 GeV and saturate to the value of 15\%. Experimental results from AGS~\cite{AKIBA1996139,PhysRevC.57.R466,20001,PhysRevC.62.024901,PhysRevC.60.064901,200053,PhysRevLett.88.102301}, SPS~\cite{PhysRevC.77.024903,PhysRevC.73.044910,PhysRevC.69.024902,PhysRevC.66.054902}, and RHIC~\cite{PhysRevC.96.044904,PhysRevC.81.024911,PhysRevC.79.034909,PhysRevLett.92.112301} of the most central collision are also shown in figure~\ref{rpi}, and are found to be consistent with the results after strong decay in the thermal model within uncertainties.

\begin{figure}
\resizebox{1\columnwidth}{!}{\includegraphics{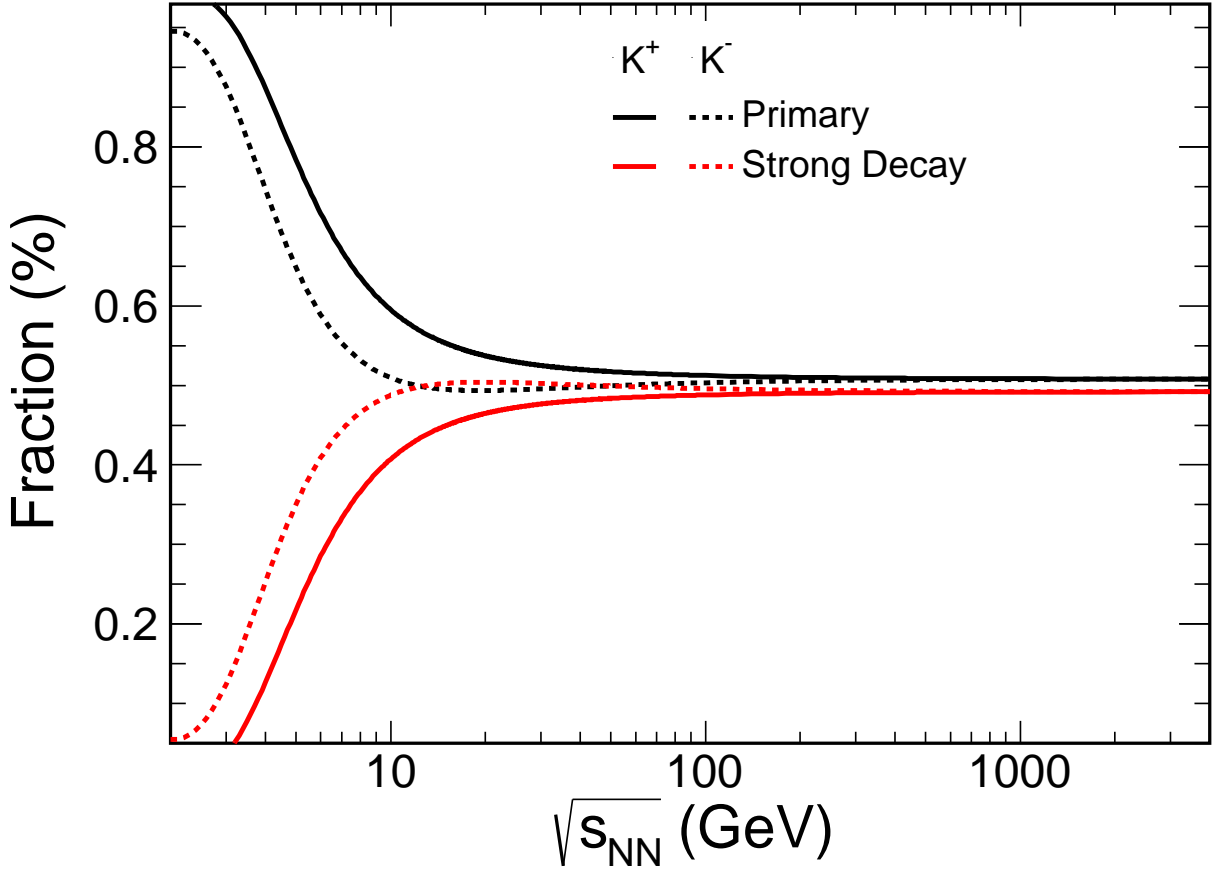}}
\caption{Energy dependence of $K^{+}$ and $K^{-}$ fractions for primary production and strong decay from high-mass resonance.}
\label{fk}
\end{figure}

\begin{figure}
\resizebox{1\columnwidth}{!}{\includegraphics{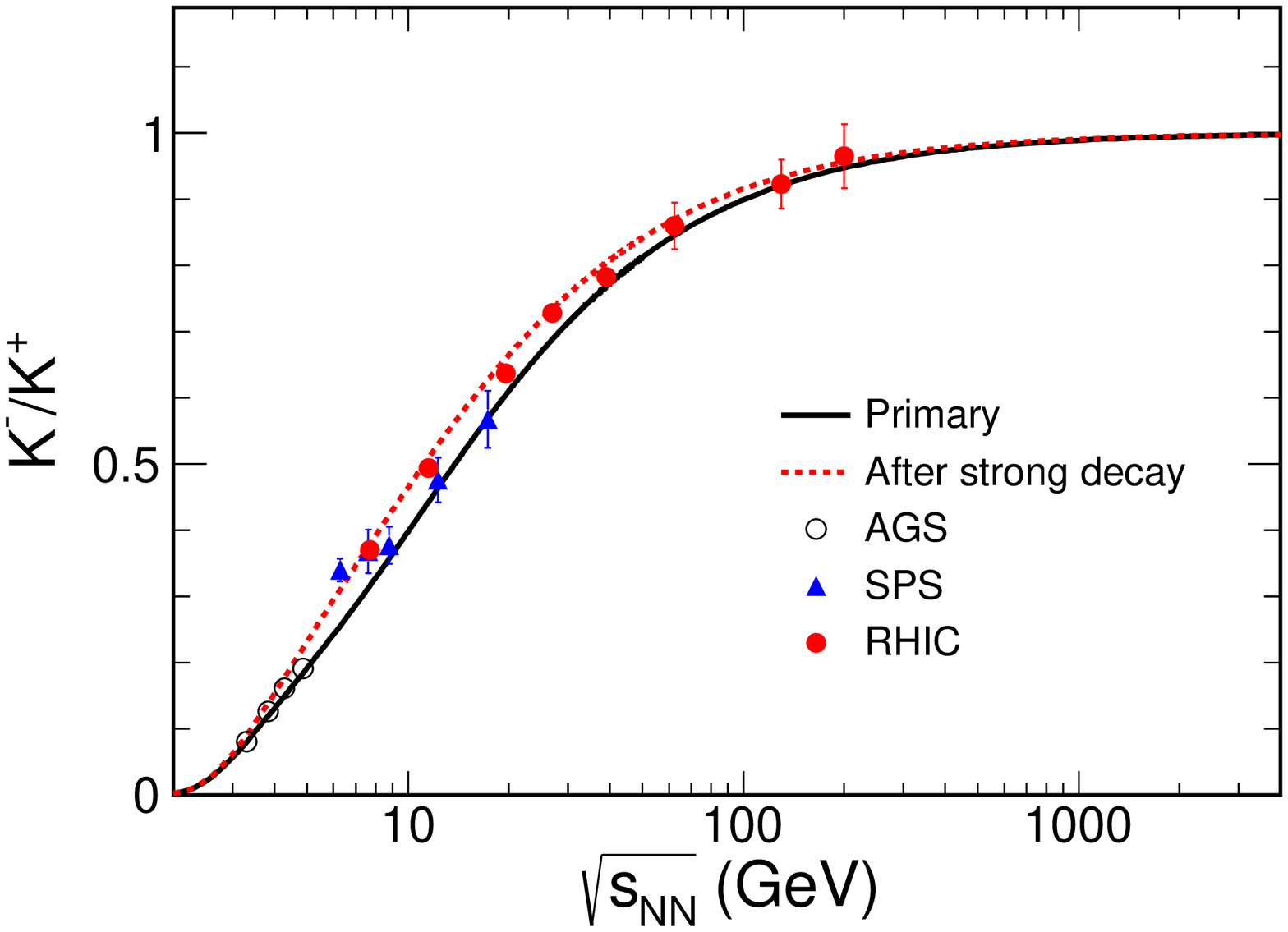}}
\caption{Energy dependence of particle ratios $K^{-}/K{+}$ from the stage at primary production, after strong decay from high-mass resonance. Experimental results from AGS~\cite{AKIBA1996139,PhysRevC.57.R466,20001,PhysRevC.62.024901,PhysRevC.60.064901,200053,PhysRevLett.88.102301}, SPS~\cite{PhysRevC.77.024903,PhysRevC.73.044910,PhysRevC.69.024902,PhysRevC.66.054902}, and RHIC~\cite{PhysRevC.96.044904,PhysRevC.81.024911,PhysRevC.79.034909,PhysRevLett.92.112301} of the most central collision are shown for comparison.}
\label{rk} 
\end{figure}

Figure~\ref{fk} shows the energy dependence of $K^{+}$ and $K^{-}$ fractions for primary production and strong decay. The weak decay channel for $K^{+}$ and $K^{-}$ is $\Omega^{-}(\bar{\Omega}^{+})\rightarrow \Lambda(\bar{\Lambda}+K^{\pm})$, which could be negligible due to the low multiplicity of $\Omega$. Energy dependence of $h_p$ and $h_s$ for $K^{\pm}$ are opposite, $h_p$ decrease and $h_s$ increase with increasing collision energy. In figure~\ref{rk}, we can find that the yield of $K^{+}$ from primary is larger than $K^{-}$, since some of the constituent $u$ quarks are from initial nucleon for $K^{+}$ but all constituent quarks ($\bar{u}$ and $s$) are from pair productions for $K^{-}$. The particle ratio $K^{-}/K^{+}$ from primary production can be written as 

\begin{equation}\label{kmkp}
\frac{h_p(K^{-})}{h_p(K^{+})}=\exp{\frac{-2\mu_S-2\mu_Q}{T_{\textrm{ch}}}},
\end{equation}
this ratio is less than unity due to $\mu_S > -\mu_Q$. At large $\snn$, $\mu_S$ and $\mu_Q$ tend to be zero, the ratio is approaching unity. Strong decay for kaon is mainly from hidden strange mesons and open strange meson. The first kind of meson decay gives the same contribution to the yields of $K^{+}$ and $K^{-}$, which will dilute the $K^{-}/K^{+}$ ratio. The second will remain unchanged in the ratio because strangeness is conserved in strong interaction. These two effects will slightly enhance the $K^{-}/K^{+}$ ratio after strong decay.

\begin{figure}
\resizebox{1\columnwidth}{!}{\includegraphics{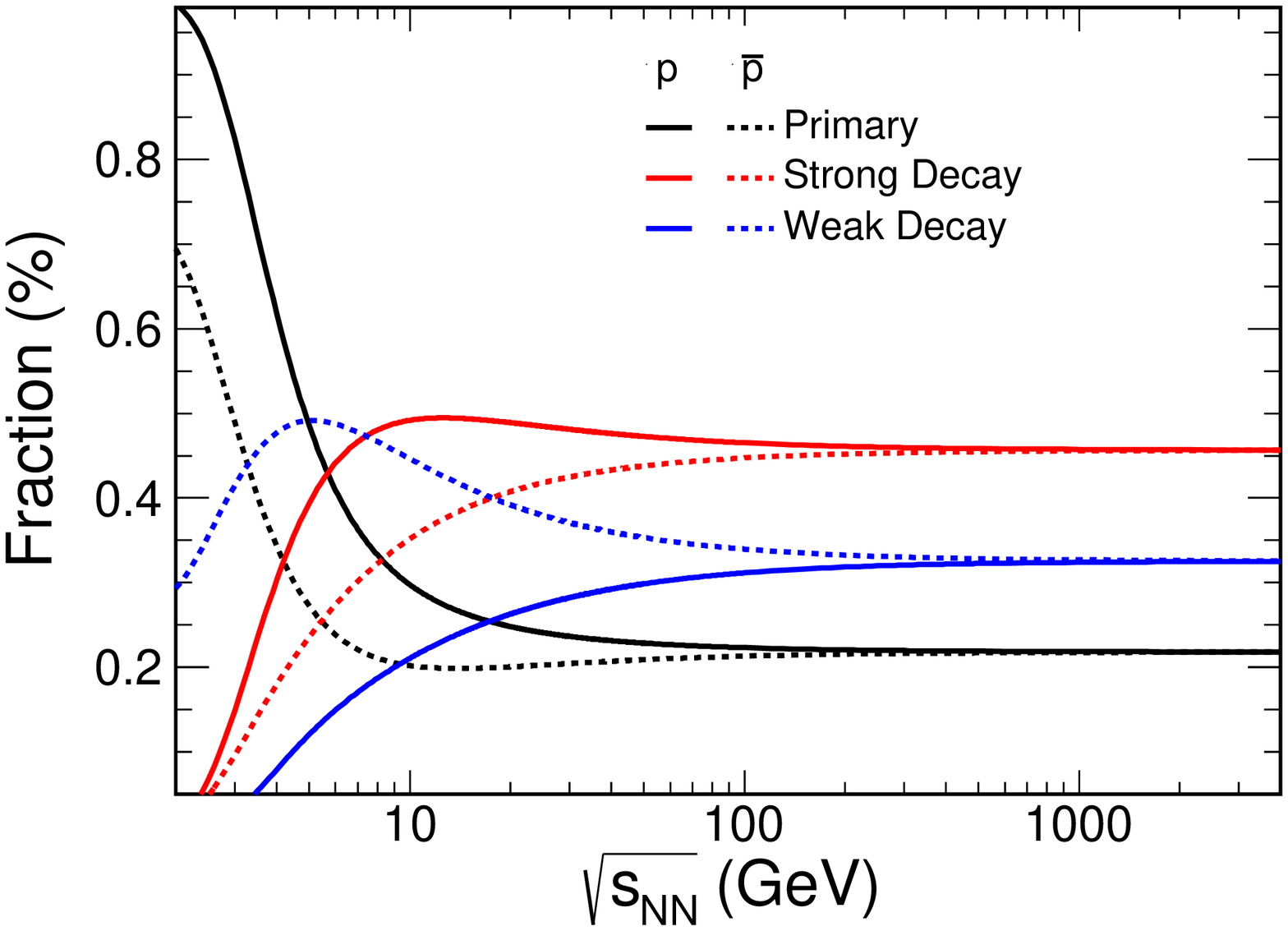}}
\caption{Energy dependence of $p$ and $\bar{p}$ fractions for primary production, strong decay from high-mass resonance, and weak decay from heavy flavor hadrons.}
\label{fp} 
\end{figure}

\begin{figure}
\resizebox{1\columnwidth}{!}{\includegraphics{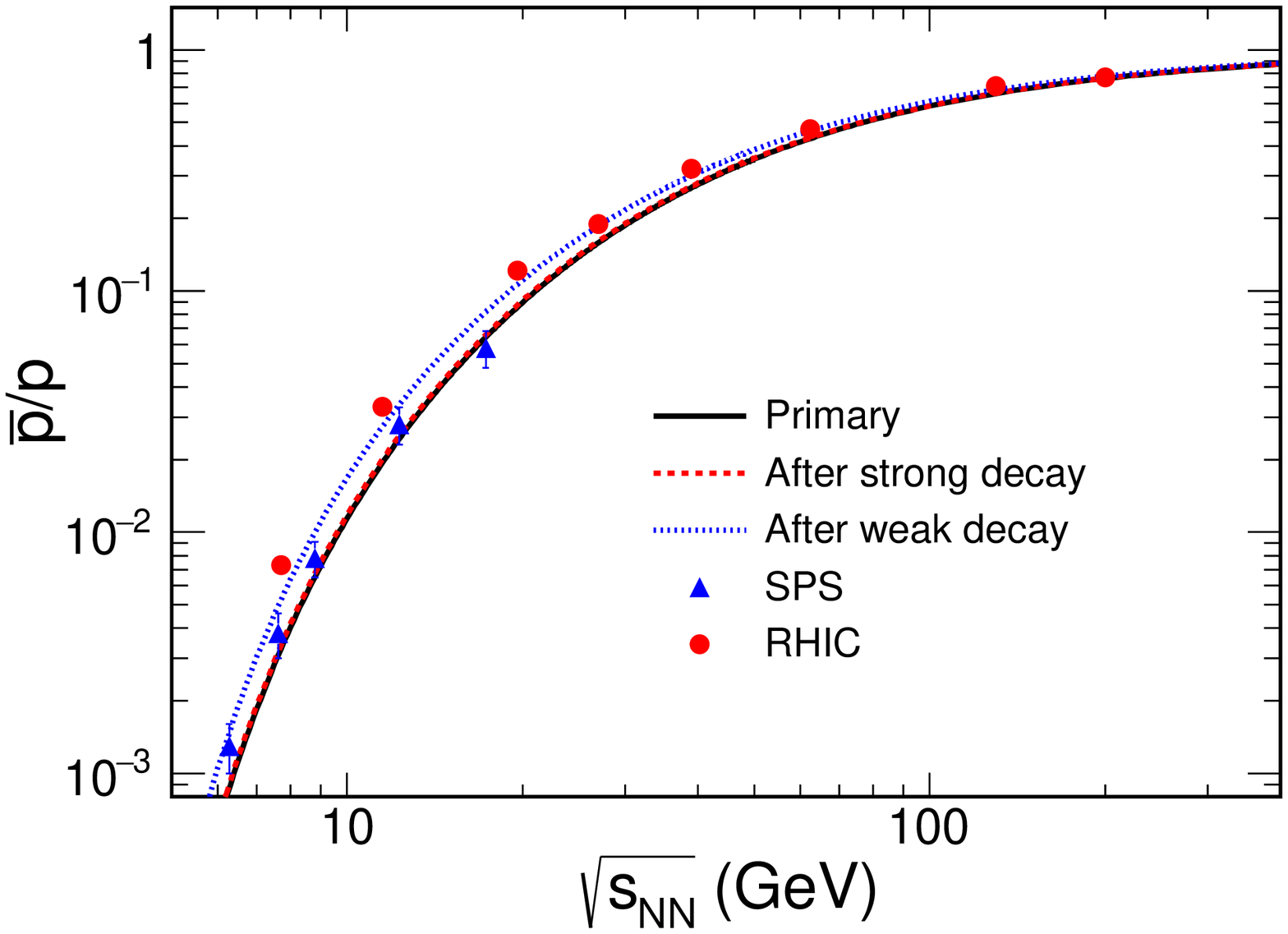}}
\caption{Energy dependence of particle ratios $\bar{p}/p$ from the stage at primary production, after strong decay from high-mass resonance. Experimental results from SPS~\cite{PhysRevC.77.024903,PhysRevC.73.044910,PhysRevC.69.024902,PhysRevC.66.054902} and RHIC~\cite{PhysRevC.96.044904,PhysRevC.81.024911,PhysRevC.79.034909,PhysRevLett.92.112301} of the most central collision are shown for comparison.}
\label{rp}
\end{figure}

Energy dependence of $p$ and $\bar{p}$ factions for primary production, strong decay from high-mass resonance, and weak decay from heavy flavor hadrons are shown in figure~\ref{fp}. The $h_s$ for $p(\bar{p})$ decrease with increasing energy and saturate to the value of 22\%. Strong decay for (anti-)proton is mainly from $\Delta$ resonance, which increase with energy and reach the maximum around $\snn = $10 GeV for $p$ and saturate to the value of 45\%. The energy dependence of $h_w$ for $p$ and $\bar{p}$ are different. $h_w(\bar{p})$ increase with energy and reach maximum around $\snn =$ 6 GeV, while $h_w(p)$ increase with energy. Both of them saturate at higher energy to the value of 33\%. In figure~\ref{rp}, particle ratios $\bar{p}/p$ of primary production, after strong decay, and after weak decay are shown. The ratio $h_p(\bar{p})/h_p(p)$ can be written as

\begin{equation}\label{pmpp}
\frac{h_p(\bar{p})}{h_p(p)}=\exp{\frac{-2\mu_B-2\mu_Q}{T_{\textrm{ch}}}}\approx \exp{-\frac{2\mu_B}{T_{\textrm{ch}}}} < 1
\end{equation}

The particle ratio $\bar{p}/p$ of $h_p+h_s$ is almost the same as the ratio of primary production due to the fact that strangeness is conserved in strong interaction and the little contribution from strange resonance baryon decay to proton.

The $p(\bar{p})$ of weak decay are from the channels below 

\begin{eqnarray}
\Lambda(\bar{\Lambda})&\rightarrow &p(\bar{p})+\pi^{\mp}\nonumber\quad\quad\textrm{B.R.}=63.9\%\\
\Sigma^{+}(\bar{\Sigma}^{-})&\rightarrow &p(\bar{p})+\pi^{0}\nonumber\quad\quad\textrm{B.R.}=51.6 \%\\
\Xi^0(\bar{\Xi}^0)&\rightarrow &\Lambda(\bar{\Lambda})+\pi^{0}\rightarrow p(\bar{p})+\pi^{0}+\pi^{\mp}\nonumber\quad\textrm{B.R.}=63.6\%\\
\Xi^-(\bar{\Xi}^+)&\rightarrow &\Lambda(\bar{\Lambda})+\pi^{-}\rightarrow p(\bar{p})+\pi^{-}+\pi^{\mp}\nonumber\quad\textrm{B.R.}=63.8\%
\end{eqnarray}
The ratios of $p(\bar{p})$ in weak decay can be calculated by
\begin{eqnarray}
\frac{h_w(\bar{p}\leftarrow\bar{\Lambda})}{h_w(p\leftarrow\Lambda)}&=&\exp{\frac{2\mu_S-2\mu_B}{T_{\textrm{ch}}}}>\frac{h_p(\bar{p})}{h_p(p)}\\
\frac{h_w(\bar{p}\leftarrow\bar{\Sigma}^{-})}{h_w(p\leftarrow\Sigma^{+})}&=&\exp{\frac{-2\mu_Q-2\mu_S-2\mu_B}{T_{\textrm{ch}}}}>\frac{h_p(\bar{p})}{h_p(p)}\\
\frac{h_w(\bar{p}\leftarrow\bar{\Xi}^0)}{h_w(p\leftarrow\Xi^0)}&=&\exp{\frac{4\mu_S-2\mu_B}{T_{\textrm{ch}}}}>\frac{h_p(\bar{p})}{h_p(p)}\\
\frac{h_w(\bar{p}\leftarrow\bar{\Xi}^+)}{h_w(p\leftarrow\Xi^-)}&=&\exp{\frac{2\mu_Q+4\mu_S-2\mu_B}{T_{\textrm{ch}}}}>\frac{h_p(\bar{p})}{h_p(p)},
\end{eqnarray}
so particle ratio $\bar{p}/p$ will be enhanced by weak decay. Experimental results from SPS~\cite{PhysRevC.77.024903,PhysRevC.73.044910,PhysRevC.69.024902,PhysRevC.66.054902}, which are corrected by weak decay, and RHIC~\cite{PhysRevC.96.044904,PhysRevC.81.024911,PhysRevC.79.034909,PhysRevLett.92.112301} of inclusive production at the most central collision are also shown in figure~\ref{rp} and are found to be consistent with the corresponding thermal model lines.

\begin{figure}
\resizebox{1\columnwidth}{!}{\includegraphics{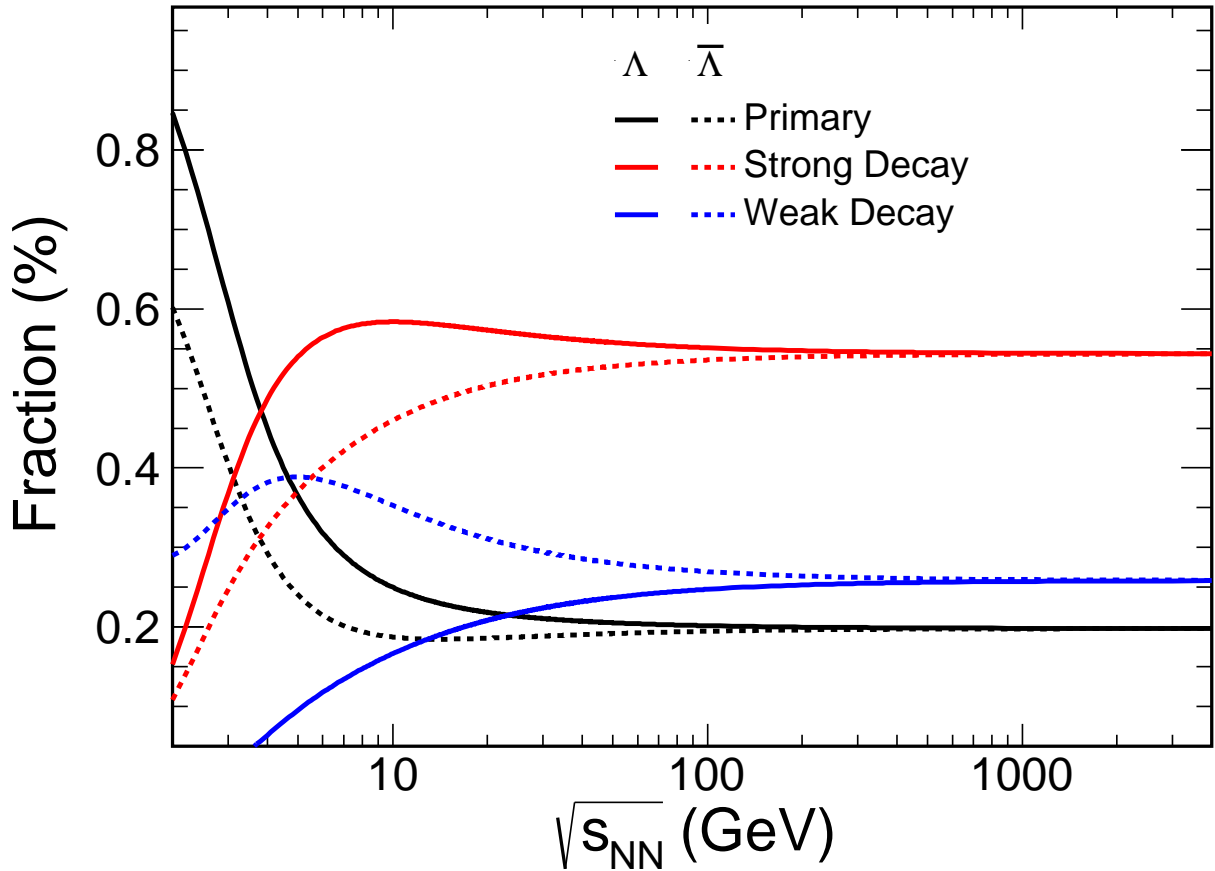}}
\caption{Energy dependence of $\Lambda$ and $\bar{\Lambda}$ fractions for primary production, strong decay from high-mass resonance, and weak decay from heavy flavor hadrons.}
\label{fl}
\end{figure}

Figure~\ref{fl} shows energy dependence of $\Lambda$ and $\bar{\Lambda}$ fractions for primary production, strong decay from high-mass resonance, and weak decay from heavy flavor hadrons. The behavior is similar to that of the (anti-)proton. The $h_s$ decrease with increasing energy and saturate to one fifth. The strong decay increases with energy and reaches the maximum around $\snn = $10 GeV and saturates to the value of 55\%. The energy dependences of $h_w$ for $\Lambda$ and $\bar{\Lambda}$ are different. $h_w(\bar{\Lambda})$ increase with energy and reach maximum around $\snn =$ 5 GeV, while $h_w(p)$ increases with energy. Both of them saturate at higher energy to the value of one fourth.

\begin{figure}
\resizebox{1\columnwidth}{!}{\includegraphics{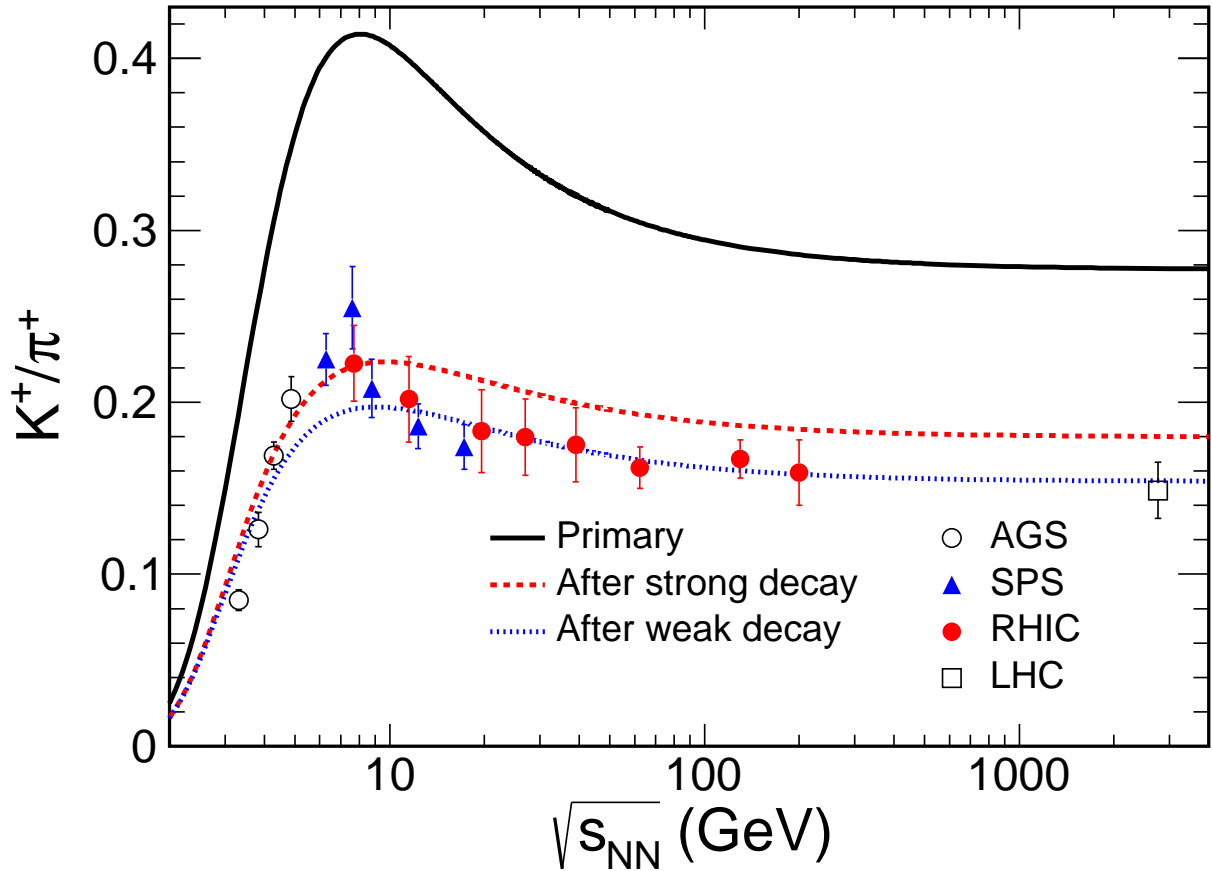}}
\caption{Energy dependence of $K^{+}/\pi^{+}$ ratio for primary production, strong decay from high-mass resonance, and weak decay from heavy flavor hadrons. Experimental results from AGS~\cite{AKIBA1996139,PhysRevC.57.R466,20001,PhysRevC.62.024901,PhysRevC.60.064901,200053,PhysRevLett.88.102301}, SPS~\cite{PhysRevC.77.024903,PhysRevC.73.044910,PhysRevC.69.024902,PhysRevC.66.054902}, and RHIC~\cite{PhysRevC.96.044904,PhysRevC.81.024911,PhysRevC.79.034909,PhysRevLett.92.112301} of the most central collision are shown for comparison.}
\label{rkpi-p}
\end{figure}

\begin{figure}
\resizebox{1\columnwidth}{!}{\includegraphics{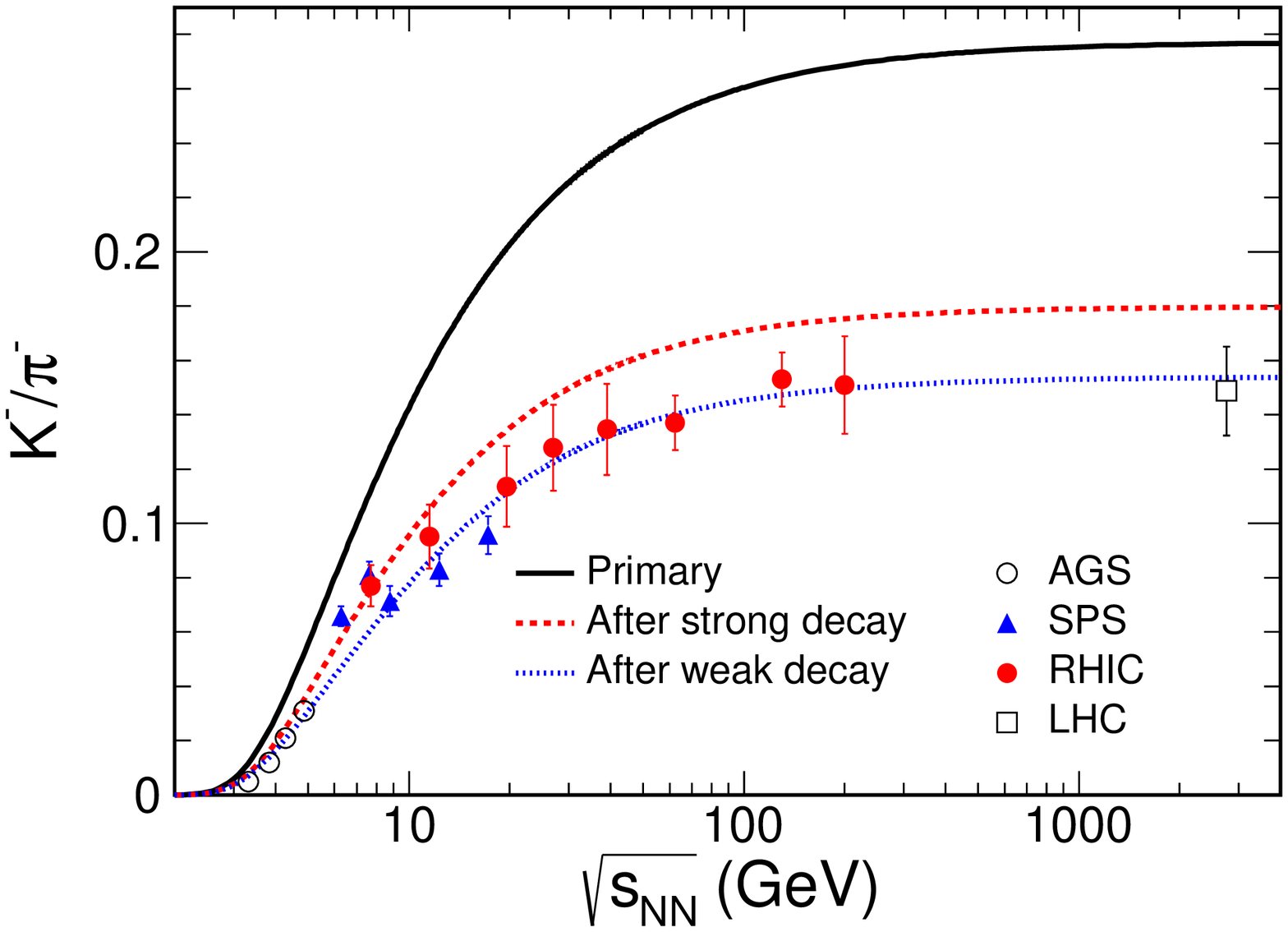}}
\caption{Energy dependence of $K^{-}/\pi^{-}$ ratio for primary production, strong decay from high-mass resonance, and weak decay from heavy flavor hadrons. Experimental results from AGS~\cite{AKIBA1996139,PhysRevC.57.R466,20001,PhysRevC.62.024901,PhysRevC.60.064901,200053,PhysRevLett.88.102301}, SPS~\cite{PhysRevC.77.024903,PhysRevC.73.044910,PhysRevC.69.024902,PhysRevC.66.054902}, and RHIC~\cite{PhysRevC.96.044904,PhysRevC.81.024911,PhysRevC.79.034909,PhysRevLett.92.112301} of the most central collision are shown for comparison.}
\label{rkpi-m}
\end{figure}

We cannot exhaust all the physical observables, therefore we show the energy dependence of particle ratio $K^{+}/\pi^{+}$ and $K^{-}/\pi^{-}$ with experimental results from AGS~\cite{AKIBA1996139,PhysRevC.57.R466,20001,PhysRevC.62.024901,PhysRevC.60.064901,200053,PhysRevLett.88.102301}, SPS~\cite{PhysRevC.77.024903,PhysRevC.73.044910,PhysRevC.69.024902,PhysRevC.66.054902}, and RHIC~\cite{PhysRevC.96.044904,PhysRevC.81.024911,PhysRevC.79.034909,PhysRevLett.92.112301} of the most central collision in figure~\ref{rkpi-p} and ~\ref{rkpi-m} for example. The $K^{+}/\pi^{+}$ ratio is roughly proportional to the total strangeness to entropy ratio, which is assumed to be preserved from the early stage until freeze-out~\cite{PhysRevC.77.024903}. The peak position (usually called the ``horn'') of the $K^{+}/\pi^{+}$ ratio in the energy dependence has been considered as an indication of QGP formation. It can be found that the horn does not change significantly after strong and weak decay but with diluting effect. $K^{-}/\pi^{-}$ ratio increases with $\snn$, corresponding to decreasing $\mu_S$ on $\snn$.

In summary, we concentrated on the use of the statistical thermal model (THERMUS) to understand the effects of strong and weak decay for different particle species which are difficult to measure in heavy-ion collision. The fractions of primary production for final hadrons decrease with increasing collision energy and somehow saturates near $\snn = $10 GeV. The appearance of this behavior can be related to specific dependence of $T_\textrm{ch}$ on the collision energy. At low energy, most of the hadrons are from primary production, while the decay components will dominate at high energy. The saturation of the primary production fraction on collision energy indicates the limitation chemical freeze-out temperature in hadronic interactions. The position of this saturation for some hadrons deviate $\snn = $10 GeV is due to the contribution of quarks that are present in the colliding particles or target and projectile. The fraction of strong decay for final hadrons increases with increasing collision energy and somehow saturates at higher collision energy, which might be related to the dependence of chemical potential ($\mu_B$ and $\mu_S$) on collision energy. The production of resonance is suppressed at large $\mu_B$ or $\mu_S$, i.e. lower collision energy but enhanced at higher collision energy. Weak decay fractions for hadron and anti-hadrons have different behavior, which may be due to the energy dependence of baryon density. The energy dependence of $h_s$ and $h_w$ will show different behavior as that of primary production at chemical freeze-out. The physical observables based on primary production hadrons are the real QCD phase diagram signals we care about. The isospin effect $\mu_Q/T_{\textrm{ch}}$ extracted from $\pi^{+}/\pi^{-}$ ratios in the experiment with the ratios corrected by weak decay are smaller than the real effect. The $K^{-}/K^{+}$ is enhanced after strong decay, while $\bar{p}/p$ does not change after strong decay and enhances after weak decay. For example, the position of the horn extracted from $K^{+}/\pi^{+}$ ratio does not vary after strong or weak decay. In this paper, we did not consider the $p_T$ or rapidity dependent of $h_s$ and $h_w$, which could show different behavior. As we know, the decay effect is dominant at low $p_T$, while the high $p_T$ particles are mainly from primary production. For future study, an extended thermal statistical model, which contains the phase space information of produced particles, should be utilized.  

We thank the fruitful discussion with Dr. Nu Xu. This work is supported in part by the MoST of China 973-Project No. 2015CB856901, the National Natural Science Foundation of China under Grants No. 11405070, 11575069, 11828501, 11890711, 11861131009, Fundamental Research Funds for the Central Universities NO. CCNU19QN054, and CCNU-QLPL Innovation Fund (Grant No. QLPL201801).
\bibliographystyle{unsrt}
\bibliography{decay}

\providecommand{\noopsort}[1]{}\providecommand{\singleletter}[1]{#1}%
\begin{thebibliography}{10}

\bibitem{ADAMS2005102}
J.~Adams et~al.
\newblock {\em Nucl. Phys.}, A757(1):102 -- 183, 2005.

\bibitem{Stephanov}
M.~A. Stephanov.
\newblock {\em Prog. Theor. Phys. Suppl.}, 153:139, 2004.

\bibitem{MOHANTY2009899c}
B.~Mohanty.
\newblock {\em Nucl. Phys.}, A830(1):899c -- 907c, 2009.

\bibitem{LUO201675}
Xiaofeng Luo.
\newblock {\em Nuclear Physics A}, 956:75, 2016.

\bibitem{Vogt}
R.~Vogt.
\newblock {\em Ultrarelativistic Heavy-Ion Collisions}.
\newblock Elsevier Science Ltd, 2007.

\bibitem{PhysRevC.65.027901}
J.~Cleymans, B.~K\"ampfer, and S.~Wheaton.
\newblock {\em Phys. Rev. C}, 65:027901, 2002.

\bibitem{PhysRevC.73.044905}
F.~Becattini, J.~Manninen, and M.~Ga\ifmmode~\acute{z}\else \'{z}\fi{}dzicki.
\newblock {\em Phys. Rev. C}, 73:044905, 2006.

\bibitem{ANDRONIC2007334}
A.~Andronic, P.~Braun-Munzinger, K.~Redlich, and J.~Stachel.
\newblock {\em Nucl. Phys.}, A789(1):334 -- 356, 2007.

\bibitem{PhysRevC.73.034905}
J.~Cleymans, H.~Oeschler, K.~Redlich, and S.~Wheaton.
\newblock {\em Phys. Rev. C}, 73:034905, 2006.

\bibitem{PhysRevC.88.044910}
B.~Abelev et~al.
\newblock {\em Phys. Rev. C}, 88:044910, 2013.

\bibitem{PhysRevLett.105.022302}
M.~M. Aggarwal et~al.
\newblock {\em Phys. Rev. Lett.}, 105:022302, 2010.

\bibitem{PhysRevLett.112.032302}
L.~Adamczyk et~al.
\newblock {\em Phys. Rev. Lett.}, 112:032302, 2014.

\bibitem{PhysRevLett.113.092301}
L.~Adamczyk et~al.
\newblock {\em Phys. Rev. Lett.}, 113:092301, Aug 2014.

\bibitem{2018551}
L.~Adamczyk et~al.
\newblock {\em Phys. Lett. B}, 785:551 -- 560, 2018.

\bibitem{Luo:2015ewa}
Xiaofeng Luo.
\newblock {\em PoS}, CPOD2014:019, 2015.

\bibitem{Luo:2017faz}
Xiaofeng Luo and Nu~Xu.
\newblock {\em Nucl. Sci. Tech.}, 28(8):112, 2017.

\bibitem{PhysRevLett.112.162301}
L.~Adamczyk et~al.
\newblock {\em Phys. Rev. Lett.}, 112:162301, 2014.

\bibitem{YU2017788}
N.~Yu.
\newblock {\em Nucl. Phys.}, A967:788--791, 2017.

\bibitem{Yu:2018kvh}
N.~Yu, D.~W. Zhang, and X.~F. Luo.
\newblock arXiv:nucl-th/1812.04291, 2018.

\bibitem{WHEATON200984}
S.~Wheaton, J.~Cleymans, and M.~Hauer.
\newblock {\em Comp. Phys. Comm.}, 180(1):84 -- 106, 2009.

\bibitem{Andronic:2017pug}
A.~Andronic, P.~Braun-Munzinger, K.~Redlich, and J.~Stachel.
\newblock {\em Nature}, 561(7723):321--330, 2018.

\bibitem{AKIBA1996139}
Y.~Akiba et~al.
\newblock {\em Nucl. Phys.}, A610:139 -- 152, 1996.

\bibitem{PhysRevC.57.R466}
L.~Ahle et~al.
\newblock {\em Phys. Rev. C}, 57:R466--R470, 1998.

\bibitem{20001}
L.~Ahle et~al.
\newblock {\em Phys. Lett. B}, 476(1):1 -- 8, 2000.

\bibitem{PhysRevC.62.024901}
J.~Barrette et~al.
\newblock {\em Phys. Rev. C}, 62:024901, 2000.

\bibitem{PhysRevC.60.064901}
L.~Ahle et~al.
\newblock {\em Phys. Rev. C}, 60:064901, 1999.

\bibitem{200053}
L~Ahle et~al.
\newblock {\em Phys. Lett. B}, 490(1):53 -- 60, 2000.

\bibitem{PhysRevLett.88.102301}
J.~L. Klay et~al.
\newblock {\em Phys. Rev. Lett.}, 88:102301, 2002.

\bibitem{PhysRevC.77.024903}
C.~Alt et~al.
\newblock {\em Phys. Rev. C}, 77:024903, 2008.

\bibitem{PhysRevC.73.044910}
C.~Alt et~al.
\newblock {\em Phys. Rev. C}, 73:044910, 2006.

\bibitem{PhysRevC.69.024902}
T.~Anticic et~al.
\newblock {\em Phys. Rev. C}, 69:024902, 2004.

\bibitem{PhysRevC.66.054902}
S.~V. Afanasiev et~al.
\newblock {\em Phys. Rev. C}, 66:054902, 2002.

\bibitem{PhysRevC.96.044904}
L.~Adamczyk et~al.
\newblock {\em Phys. Rev. C}, 96:044904, 2017.

\bibitem{PhysRevC.81.024911}
B.~I. Abelev et~al.
\newblock {\em Phys. Rev. C}, 81:024911, 2010.

\bibitem{PhysRevC.79.034909}
B.~I. Abelev et~al.
\newblock {\em Phys. Rev. C}, 79:034909, 2009.

\bibitem{PhysRevLett.92.112301}
J.~Adams et~al.
\newblock {\em Phys. Rev. Lett.}, 92:112301, 2004.

\end{thebibliography}


\end{document}